\documentstyle[epsf,amssymb,twocolumn,prd,aps,floats]{revtex}
\begin{document}

\def\lsim{\; \raise0.3ex\hbox{$<$\kern-0.75em
      \raise-1.1ex\hbox{$\sim$}}\; }
\def\gsim{\; \raise0.3ex\hbox{$>$\kern-0.75em
      \raise-1.1ex\hbox{$\sim$}}\; }
\date{August 27, 2001}

%
%
\newcommand{\nc}{\newcommand}
\newcommand{\comment}[1]{}

\nc{\bea}{\begin{eqnarray}}
\nc{\eea}{\end{eqnarray}}
\nc{\beq}{\begin{equation}}
\nc{\eeq}{\end{equation}}
\nc{\bi}{\begin{itemize}}
\nc{\ei}{\end{itemize}}
\nc{\la}[1]{\label{#1}}
\nc{\half}{\frac{1}{2}}
\nc{\fsky}{f_{\rm sky}}
\nc{\fwhm}{\theta_{\rm fwhm}}
\nc{\fwhmc}{\theta_{{\rm fwhm},c}}
\nc{\nadi}{n_{\rm adi}}
\nc{\niso}{n_{\rm iso}}
\nc{\R}{{\cal{R}}}
\nc{\GeV}{\mbox{ GeV}}
\nc{\MeV}{\mbox{ MeV}}
\nc{\keV}{\mbox{ keV}}
\nc{\etal}{{\it et al.}}
\nc{\x}[1]{}
%
%

\nc{\AJ}[3]{{Astron.~J.\ }{{\bf #1}{, #2}{ (#3)}}}
\nc{\anap}[3]{{Astron.\ Astrophys.\ }{{\bf #1}{, #2}{ (#3)}}}
\nc{\ApJ}[3]{{Astrophys.~J.\ }{{\bf #1}{, #2}{ (#3)}}}
\nc{\apjl}[3]{{Astrophys.~J.\ Lett.\ }{{\bf #1}{, #2}{ (#3)}}}
\nc{\app}[3]{{Astropart.\ Phys.\ }{{\bf #1}{, #2}{ (#3)}}}
\nc{\araa}[3]{{Ann.\ Rev.\ Astron.\ Astrophys.\ }{{\bf #1}{, #2}{ (#3)}}}
\nc{\arns}[3]{{Ann.\ Rev.\ Nucl.\ Sci.\ }{{\bf #1}{, #2}{ (#3)}}}
\nc{\arnps}[3]{{Ann.\ Rev.\ Nucl.\ and Part.\ Sci.\ }{{\bf #1}{, #2}{ (#3)}}}
\nc{\epj}[3]{{Eur.\ Phys.\ J.\ }{{\bf #1}{, #2}{ (#3)}}}
\nc{\MNRAS}[3]{{Mon.\ Not.\ R.\ Astron.\ Soc.\ }{{\bf #1}{, #2}{ (#3)}}}
\nc{\mpl}[3]{{Mod.\ Phys.\ Lett.\ }{{\bf #1}{, #2}{ (#3)}}}
\nc{\Nat}[3]{{Nature }{{\bf #1}{, #2}{ (#3)}}}
\nc{\ncim}[3]{{Nuov.\ Cim.\ }{{\bf #1}{, #2}{ (#3)}}}
\nc{\nast}[3]{{New Astronomy }{{\bf #1}{, #2}{ (#3)}}}
\nc{\np}[3]{{Nucl.\ Phys.\ }{{\bf #1}{, #2}{ (#3)}}}
\nc{\pr}[3]{{Phys.\ Rev.\ }{{\bf #1}{, #2}{ (#3)}}}
\nc{\PRD}[3]{{Phys.\ Rev.\ D\ }{{\bf #1}{, #2}{ (#3)}}}
\nc{\PRL}[3]{{Phys.\ Rev.\ Lett.\ }{{\bf #1}{, #2}{ (#3)}}}
\nc{\PL}[3]{{Phys.\ Lett.\ }{{\bf #1}{, #2}{ (#3)}}}
\nc{\prep}[3]{{Phys.\ Rep.\ }{{\bf #1}{, #2}{ (#3)}}}
\nc{\RMP}[3]{{Rev.\ Mod.\ Phys.\ }{{\bf #1}{, #2}{ (#3)}}}
\nc{\rpp}[3]{{Rep.\ Prog.\ Phys.\ }{{\bf #1}{, #2}{ (#3)}}}
\nc{\ibid}[3]{{\it ibid.\ }{{\bf #1}{, #2}{ (#3)}}}

\wideabs{

\begin{flushright}
HIP-2001-48/TH\\
astro-ph/0108422\\
Phys.~Rev.~D 65, 0230XX
\end{flushright}


\title{Open and closed CDM isocurvature models contrasted with the CMB data}

\author{Kari Enqvist\cite{mailk}}
\address{Department of Physical Sciences, University of Helsinki,
and Helsinki Institute of Physics,\\
         P.O. Box 64, FIN-00014 University of Helsinki, Finland}

\author{Hannu Kurki-Suonio\cite{mailh} and Jussi V\"{a}liviita\cite{mailv}}
\address{Department of Physical Sciences, University of Helsinki,
         P.O. Box 64, FIN-00014 University of Helsinki, Finland}


\maketitle

\begin{abstract}
We consider pure isocurvature cold dark matter
models in the case of open and closed
universes. We allow for a large spectral tilt and scan the 6-dimensional
parameter space for the best fit to the COBE, Boomerang, and Maxima-1 data.
Taking into account constraints from large-scale structure and big bang
nucleosynthesis, we find a best fit with
$\chi^2 = 121$, which is to be
compared to $\chi^2 = 44$ of a flat adiabatic reference model.
Hence the current data strongly disfavor pure isocurvature perturbations.

\end{abstract}

}

%
%
\section{Introduction}

The recent measurements
of the cosmic micro\-wave background (CMB) temperature
fluctuations by the
Boomerang\cite{boom1,boom2} and
Maxima-1\cite{max1,max2} balloon experiments and the DASI
interferometer\cite{DASI}
have widely been regarded as
indicating that we live in a $\Omega=1$ universe. This is so because
the
first acoustic peak is found at the multipole $\ell\simeq 200$, implying a
flat universe. The firmness of such a conclusion is, however, based
on certain tacit assumptions. In particular, when fitting the acoustic peak
positions, one often assumes that the primordial perturbations are
adiabatic and that the spectrum is nearly scale invariant.

If perturbations are adiabatic, the relative abundances of particle
species are equal to their thermal equilibrium values.
This is the case in the simplest, one-field inflation models
but it is not a generic feature of inflation. More generally,
perturbations can be either adiabatic or nonadiabatic; the latter
would be perturbations in the particle number densities, or entropy
perturbations, and are called isocurvature perturbations.

Because no generally accepted theory of inflation exists, it is natural
to consider both adiabatic and isocurvature perturbations as being
equally probable. This is the generic situation when more than one field
is excited during inflation, such as is the case in double inflation
\cite{double} or in the minimally supersymmetric standard model
with flat directions \cite{johncmb}.
One should also note that
in the pre-big-bang scenario, which has been proposed as an alternative
to the inflationary universe, pre-big-bang axion field fluctuations
give rise to an isocurvature perturbation spectrum \cite{pbb}.
Purely isocurvature $\Omega=1$
perturbations are, however, not consistent \cite{Stompor,Peebles,eksv} with the
observational data, but an admixture of (uncorrelated or correlated)
adiabatic and isocurvature perturbations
cannot be ruled out \cite{eksv,gb,eks,correlated}. However, if we do
not insist on a flat universe, the situation could be different.

Recently, it was pointed out \cite{turok} that
in the general (Gaussian) case the
scalar power spectrum is a $5\times 5$ matrix $P_{ij}({\bf k})=\langle
A_i({\bf k})A_j(-{\bf k})\rangle$, where $i,j$ label one adiabatic
and four isocurvature modes [cold dark matter (CDM),
baryon, neutrino density, and
neutrino velocity] and their correlations. Here we
shall focus on a purely isocurvature primordial perturbation
in the CDM which has the power spectrum
\beq
P_S(k) = B\left(\sqrt{k^2-K}\right)^{\niso-4}~,
\label{E:jvaliviita:PS}
\eeq
where $\niso$ is the spectral index and $K = -H_0^2 (1 - \Omega)$
is the curvature. Since
in curved space the Laplacian has eigenvalues $k^2-K$ instead
of the $k^2$ of the flat case, the
spectrum (\ref{E:jvaliviita:PS}) is the simplest
generalization of the flat space spectrum $k^{\niso-4}$. 

In the flat, $\Omega = 1$ case, definition (\ref{E:jvaliviita:PS}) gives
the power law $P_S(k) \propto k^{\niso-4}$, which is a natural form
for the power spectrum, and approximates well the spectrum produced
by typical inflation models with isocurvature perturbations in the region of
interest.  
The scale-invariant spectrum has $\niso=1$. In principle, $\niso$ could
well depend on $k$; here we shall assume that it is a constant
(or varies very little) over the range of interest.
In open and closed models the spatial curvature introduces a
length scale and one expects this to be reflected in the form
of the power spectrum.
It is not obvious what would
be the most natural modification to the power law
for isocurvature models in curved space.
This question has been
studied only for specific models in the adiabatic
case\cite{curvedspace1,curvedspace2}.
Thus we stress that we are using a phenomenological power-law spectrum,
which does not necessarily follow from any  particular inflation model.
We shall return to this point later in this paper.

After the clear detection of the acoustic peak around
$\ell \simeq 200$ it became evident that the adiabatic models
fit well to the data\cite{boom1,boom2,max2,DASI,Wang,Douspis}.
However, this should not be taken as a proof that all pure
isocurvature models are ruled out.
Some unconventional
combination of cosmological parameters, e.g., $\Omega\ne 1$ and a spectrum
with a large tilt, could at least in principle
give an equally good fit as do the
adiabatic models.

Pure isocurvature models
have two well-recognized problems: excess power
at low multipoles
and a peak structure that is roughly speaking out of phase
by $\pi/2$ when compared to the adiabatic one \cite{HuWhite96}.
Since the angular power in the low multipole
region was measured quite firmly by 
the Cosmic Background Explorer
(COBE), $\chi^2$ fitting forces the
overall normalization constant in pure isocurvature models to be
smaller than in the adiabatic case, which
leads to too little power at higher multipoles.
The easiest and perhaps the only way to compensate for this is to introduce
a large spectral tilt.
Moreover, since flat adiabatic
models fit the observed peak at  $\ell \simeq 200$ well, it is obvious that
the $\ell \simeq 200$ peak falls between the first and second peaks of any
flat isocurvature model.
Accordingly, in our earlier study  \cite{eksv}, the best-fit flat
isocurvature model was found to have a large
$\chi^2 = 116$ for 30 data points and 6 parameters
whereas the best adiabatic model had
$\chi^2 = 22$.

Thus we have two possibilities for a better isocurvature model.
The first is to
lower the total energy density parameter so much that
the position of the first isocurvature peak fits to the
observed peak at $\ell \simeq 200$, which means that we have to allow for
an open
universe ($\Omega < 1$). The other possibility is
to increase the total energy
density parameter so much that
the position of the second isocurvature peak fits
the $\ell \simeq 200$ peak \cite{DurrerMelchiorri}, implying a
closed universe ($\Omega > 1$).
In this case the
first isocurvature peak at $\ell \simeq 60 \ldots 100$
should effectively disappear.
In fact, a large spectral tilt would have precisely this effect since it
would decrease
the relative power at low $\ell$.

The purpose of the present paper is to study these possibilities systematically
to find out if CDM isocurvature models are indeed completely ruled
out by the presently available CMB data.

\section{Methods and results}

In order to compare the isocurvature models with adiabatic ones we choose
one representative well-fitted adiabatic model
$(\nadi,\Omega_m,\Omega_\Lambda,\omega_b,\omega_c,\tau)
= (0.98, 0.38, 0.62, 0.021, 0.13, 0)$; cf.~\cite{boom1}.
Using the same data sets and algorithm as for
isocurvature models, we get $\chi^2 = 44$ for this adiabatic ``reference''
model. Fig.~\ref{fig:3}(b) confirms that this model fits well both the
low $\ell$ part of the angular power spectrum and the acoustic peaks. 

Our starting point for analyzing isocurvature models is a large
grid with the following free parameters:
\begin{itemize}
\item $\niso = 1.00 \ldots 7.00$ (60 values)
\item $\Omega_m = 0.06 \ldots 2.31$ (16 values)
\item $\Omega_\Lambda = -1.00 \ldots 1.10$ (14 values)
\item $\omega_b = 0.001 \ldots 0.100$ (10 values)
\item $\omega_c = 0.01 \ldots 1.60$ (15 values),
\end{itemize}
where $\Omega_m$ is the total matter density, $\Omega_\Lambda$ is the
vacuum energy density,
$\omega_b = h^2\Omega_b$ is the baryon density,
and $\omega_c = h^2\Omega_c$ is the CDM density. The sixth free
parameter is the overall normalization factor $B$
of Eq.~(\ref{E:jvaliviita:PS}).
The Hubble constant $h$ is not a free parameter, since
$h^2\Omega_m = \omega_m = \omega_b+\omega_c$.
We use a top-hat prior $h = 0.45 \ldots 0.90$
and assume $\tau = 0$ for the optical depth due to reionization.
The angular power
spectrum of all the models in the grid
was calculated by CAMB \cite{camb} assuming isocurvature CDM initial conditions.

We use the $\chi^2$ method to compare models and data, because it
allows us to quickly search a large parameter space.  This method is
approximate\cite{curvedspace2} and we do not attempt precise estimates for
cosmological parameters or confidence levels.
As will be seen, the conclusion is clear enough in ruling out
the isocurvature models
so that it is not necessary to go to a full maximum
likelihood analysis\cite{bondjaffe}.

\begin{figure}[!t]
\epsfysize=7.0cm
\epsffile{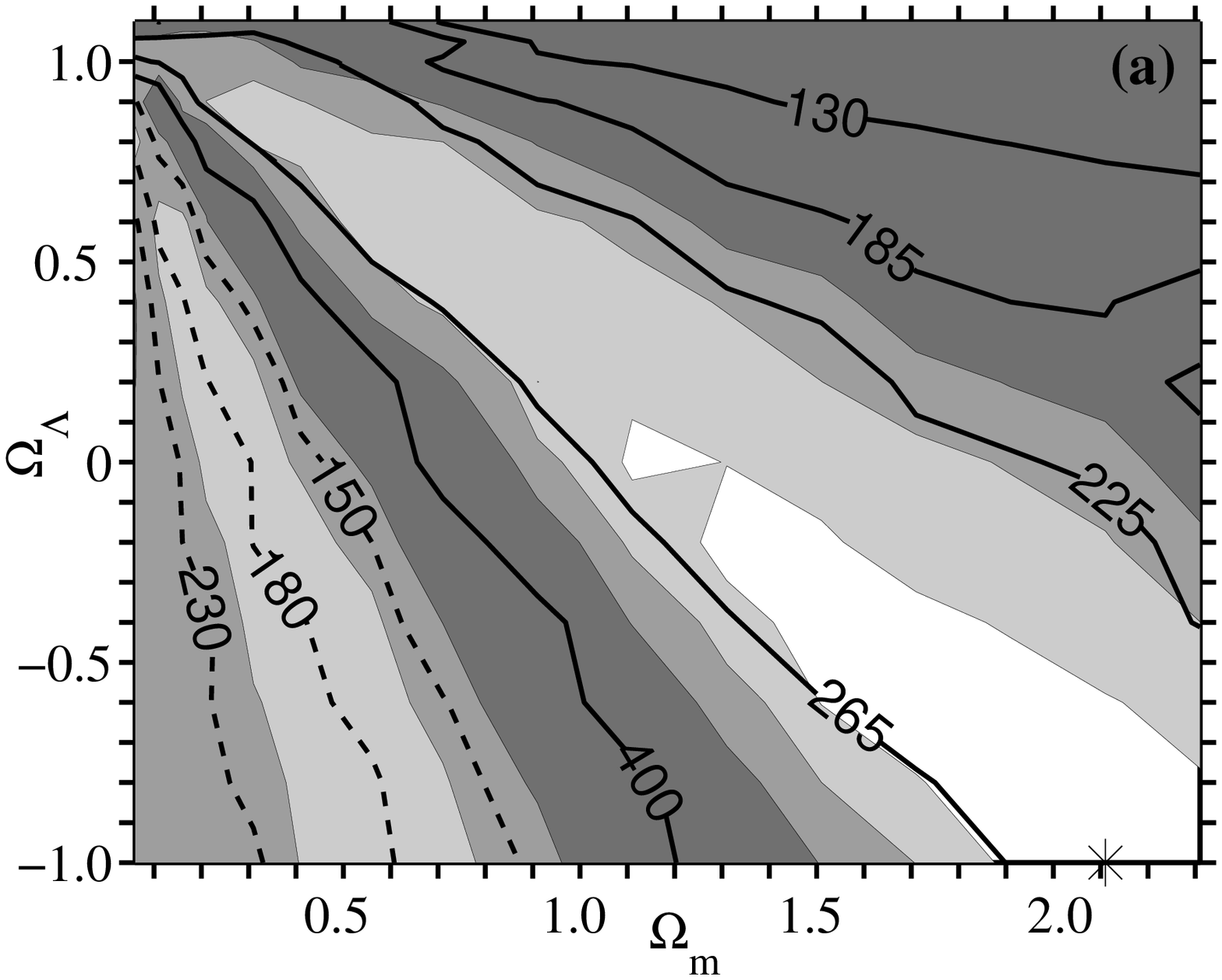}
\epsfysize=7.0cm
\epsffile{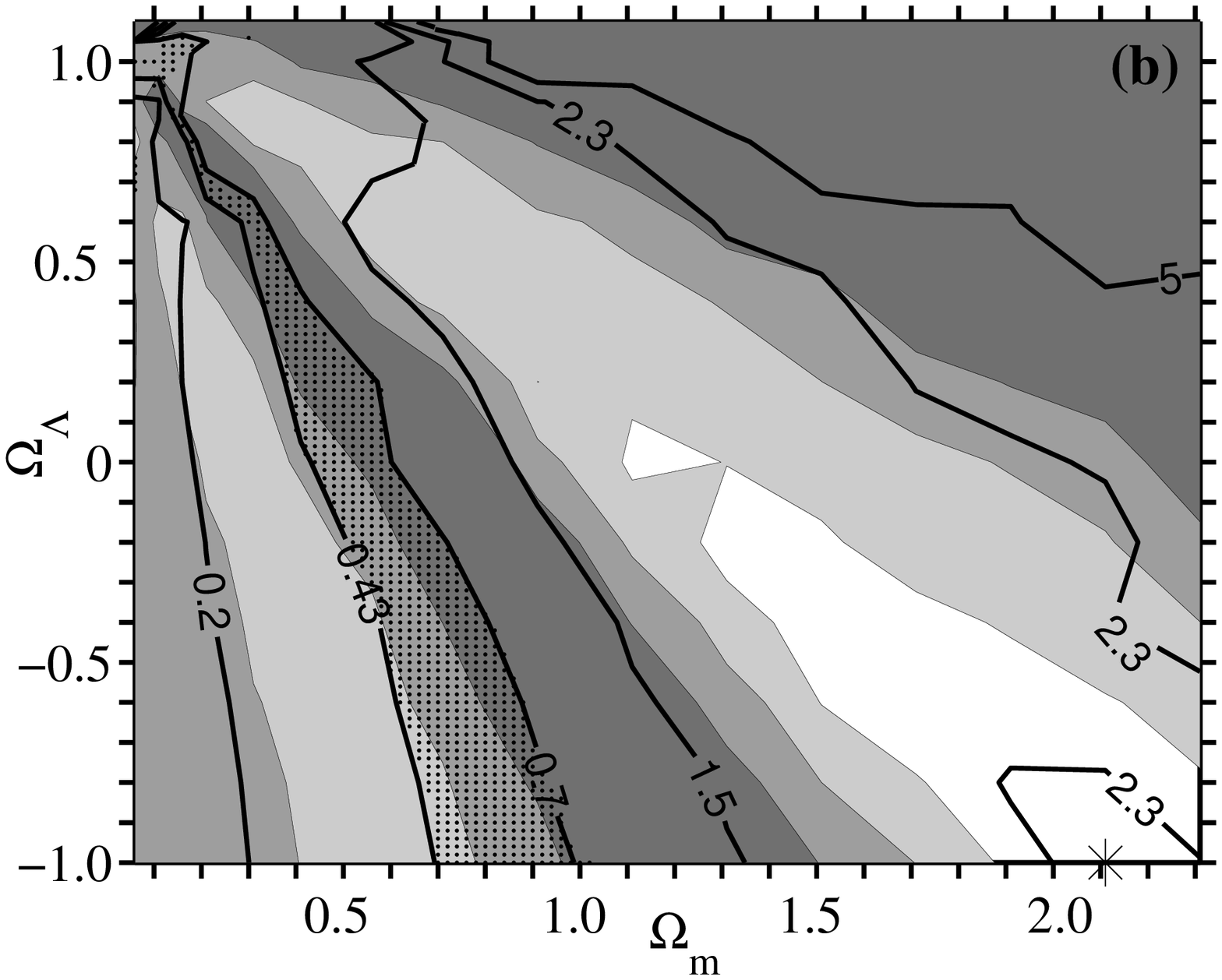}
\vspace{0.01cm}
\caption[a]{\protect
The best-$\chi^2$ contours on the $(\Omega_m,\Omega_\Lambda)$ plane.
The best fit, which has $\chi^2=80$, is indicated by an
asterisk ($\ast$) near to the lower right corner.
The contours for deviation from the best fit are as follows:
white $\Delta\chi^2 < 10$;
light gray $10 < \Delta\chi^2 < 40$;
medium gray $40 < \Delta\chi^2 < 100$;
and dark gray $\Delta\chi^2 > 100$.
(a) Dashed lines show
the position ($\ell$) of the first acoustic peak
and solid lines the second peak. 
(b) Solid lines give the values of
$\sigma_8\Omega_m^{0.56}$, and the dotted area is that allowed by the LSS
constraint $0.43 < \sigma_8\Omega_m^{0.56} < 0.70$.
} \label{fig:1}
\vspace{1.15cm}
\end{figure}

\begin{figure}[!t]
\epsfysize=7.0cm
\smallskip
\epsffile{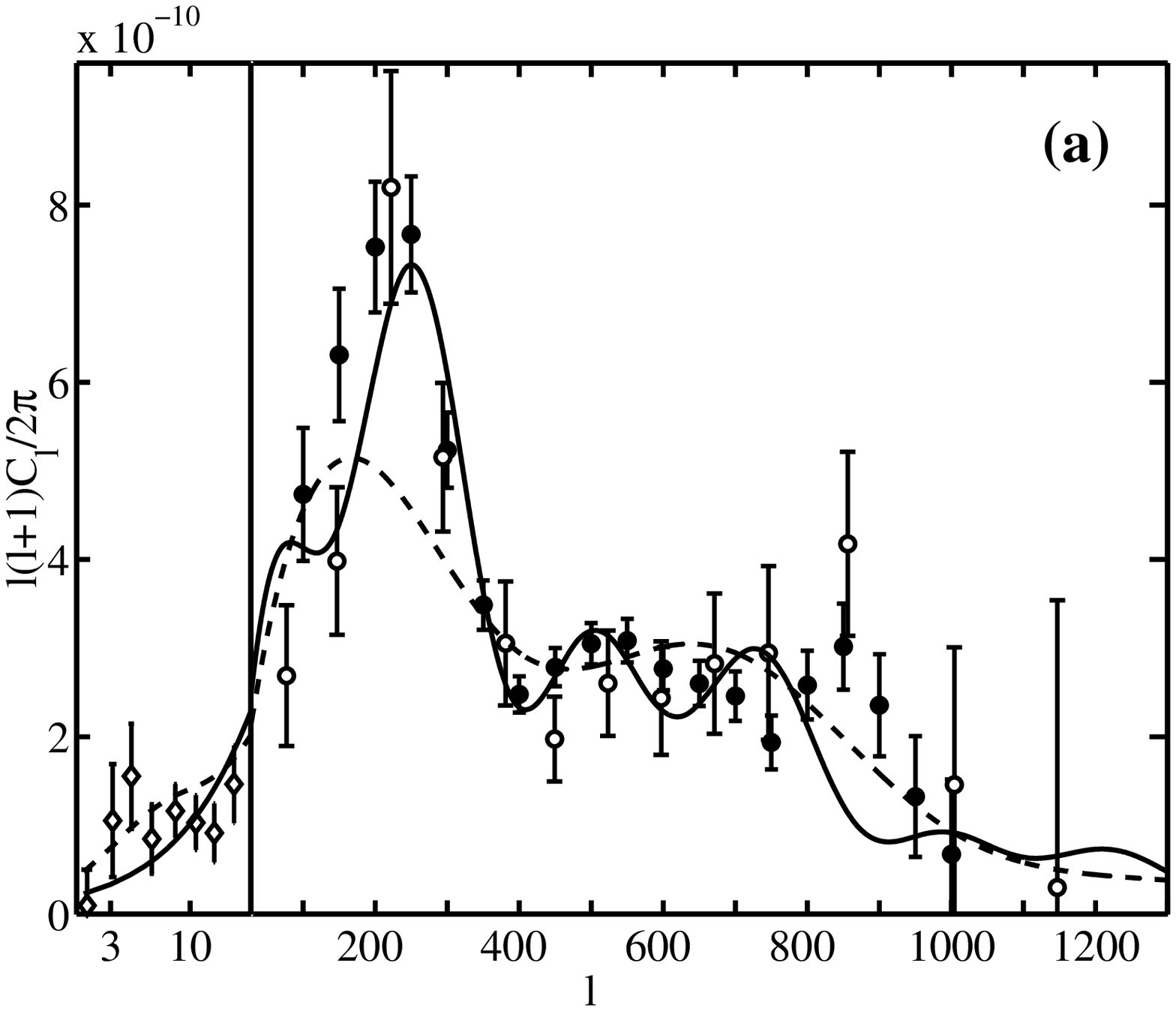}
\epsfysize=7.0cm
\epsffile{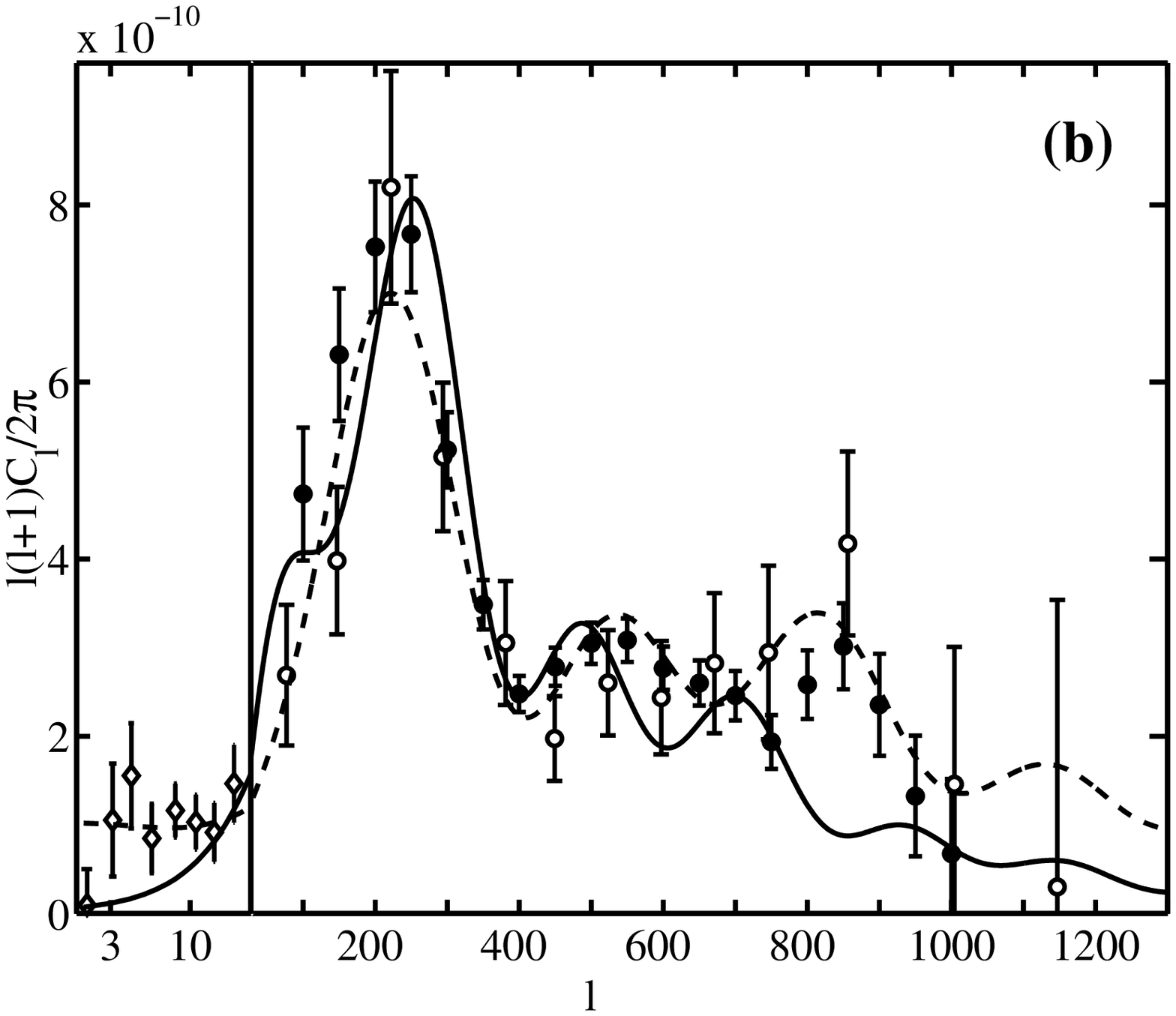}
\caption[a]{\protect
Angular power spectra for different models
along with COBE ($\diamond$), Boomerang ($\bullet$) , and Maxima-1 ($\circ$)
data.
(a) Best-fit isocurvature model
of Fig.~\ref{fig:1}
(solid line) and best-fit open model with LSS constraint
(dashed line).
(b) Best physical isocurvature
fit from the fine grid (solid line) and the adiabatic reference model
(dashed line).
Note that up to $\ell=25$ the $\ell$ axis is logarithmic.}
\label{fig:3}
\vspace{0.3cm}
\end{figure}

Using the latest Boomerang data \cite{boom1}, 
together with Maxima-1 \cite{max1} and
COBE data \cite{TegZal00}
we calculate $\chi^2$ for each model. The resulting
best-$\chi^2$ contours in the $(\Omega_m,\Omega_\Lambda)$ plane
are presented in Fig.~\ref{fig:1} by gray levels.
The best-fit model
turns out to have $\chi^2 = 80$ with
$(\niso,\Omega_m,\Omega_\Lambda,\omega_b,\omega_c)
= (2.00, 2.11, -1.00, 0.020, 1.40)$.
From Fig.~\ref{fig:1}(a) we see that
the best-fit isocurvature models lie along two bands in the
$(\Omega_m,\Omega_\Lambda)$ plane,
the left band corresponding to open universes, and the
right corresponding to closed universes.
In the best-fit models the spectral index falls
in the range $\niso = 2\ldots 3$.

A detailed examination of the various pure isocurvature models
allows us to conclude that
the main problems are the spacings of the
higher acoustic peaks and the slope in the (low $\ell$)
Sachs--Wolfe region. COBE measured a close-to-flat $C_\ell$ spectrum, but
the isocurvature models have a significant positive slope arising from the
large primordial blue spectral tilt needed to
get enough power at higher multipoles.

In the best-fit open models the prominent peak in the CMB data is fitted by the
first acoustic peak of the isocurvature model.
Fig.~\ref{fig:1}(a) shows
that in the best-fit open region the first peak lies
in the range $150 \lsim \ell \lsim 230$.
Since the data do not show a high second peak, these models need a
small baryon density $\omega_b$ to boost up the first peak and suppress the
second peak. (In the adiabatic case, adding more baryons enhances odd
acoustic peaks over even \cite{HuWhite96},
but in the isocurvature case increasing $\omega_b$ boosts even
peaks.) Actually, all
the best-fit open models have a baryon density of $\omega_b = 0.001$,
which is the smallest value in the grid. However, even assuming
such an unphysically low baryon density as
$0.0005$ only gives about half of the power needed to fit the first peak,
so not scanning below $\omega_b<0.001$ seems justified.
\begin{figure}[!t]
\epsfysize=7.0cm
\epsffile{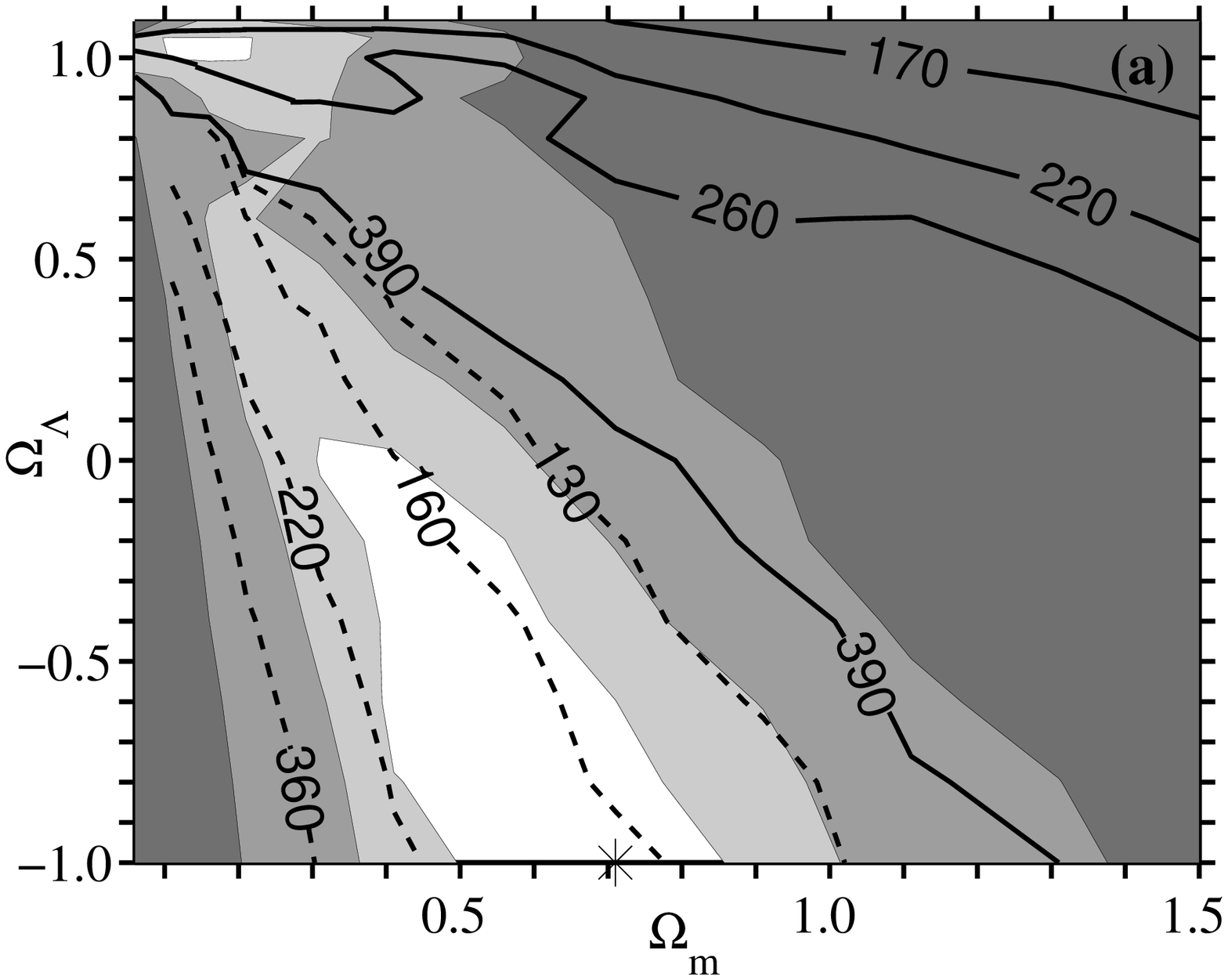}
\epsfysize=6.8cm
\epsffile{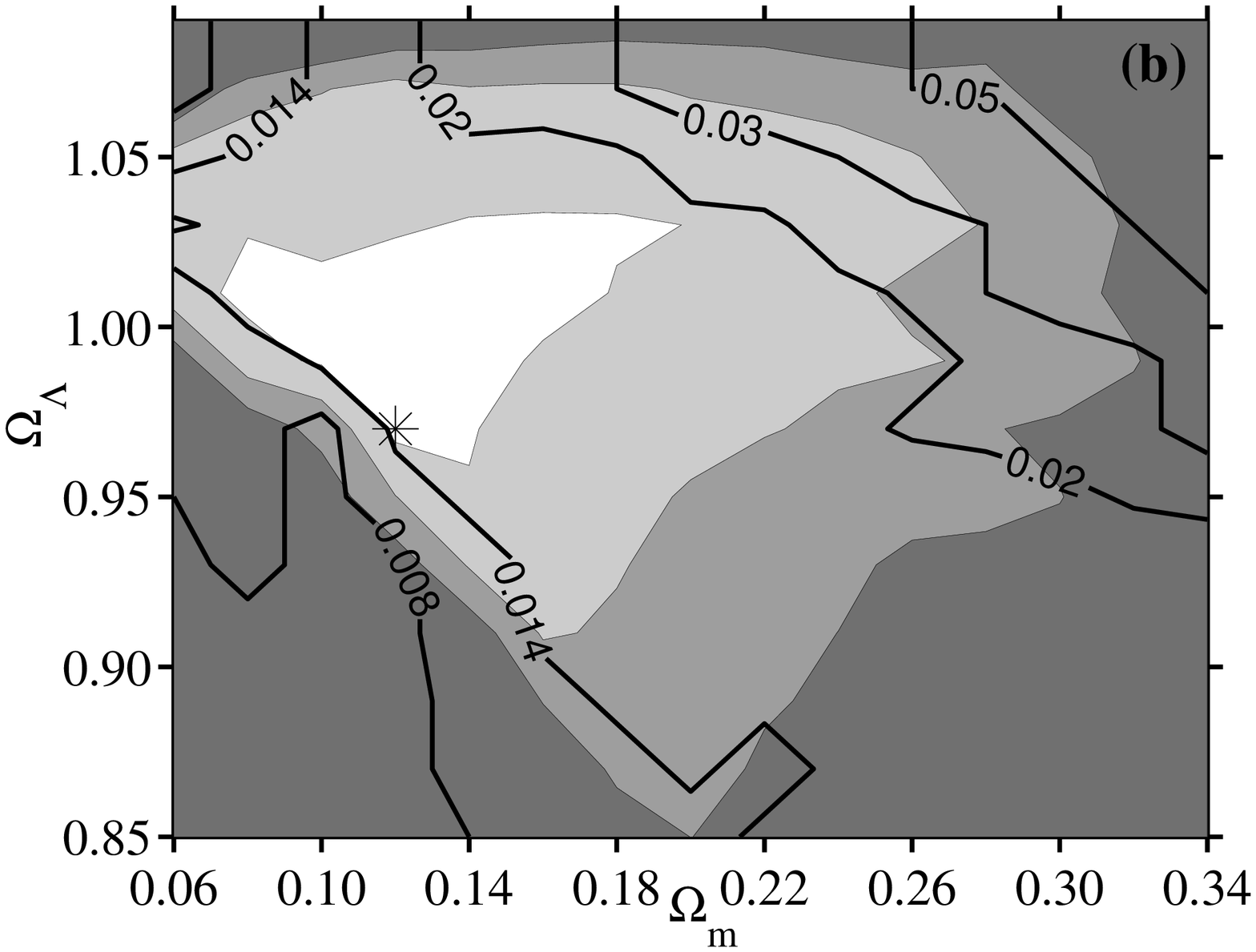}
\vspace{0.24cm}
\caption[a]{\protect
(a) As Fig.~\ref{fig:2}(a) but now
with the LSS constraint $0.43 < \sigma_8\Omega_m^{0.56} < 0.70$.
The best fit marked by an asterisk has $\chi^2=103$. The contours
for deviation from the best fit are as follows: white
$\Delta\chi^2 < 35$; light
gray $35 < \Delta\chi^2 < 140$; medium gray
$140 < \Delta\chi^2 < 350$; and dark gray $\Delta\chi^2 > 350$.
The upper left corner corresponds to the closed models where the
second acoustic peak fits the prominent peak in the $C_\ell$ data.
(b) The best-fit physical region using the fine grid.
The solid contours show the baryon density $\omega_b$.
The best-fit model has $\chi^2 = 121$ and the gray levels are
as follows:
white $\Delta\chi^2 < 6$;
light gray $6 < \Delta\chi^2 < 30$,
medium gray $30 < \Delta\chi^2 < 60$,
and dark gray $\Delta\chi^2 > 60$.
} \label{fig:2}
\end{figure}

In the best-fit closed models the $\ell \simeq 200$
peak in the CMB data is fitted by the
second isocurvature peak, which lies, according to Fig.~\ref{fig:1}(a), in
the range $225 \lsim \ell \lsim 265$.
As one might expect (see, e.g., \cite{hu}
for an adiabatic analogy),
now the ratio of the $\ell \simeq 200$
peak to the higher multipole $C_\ell$'s in the data fixes
$\omega_b$
near the value $0.02$ in the whole best-fit band.
In contrast one obtains almost no
restriction for $\omega_c$. This is consistent with
Fig.~\ref{fig:1}, where $\Omega_m$ can be seen to be able to take
almost any value, which is then compensated by $\Omega_\Lambda$ to
produce the correct peak position.

According to Fig.~\ref{fig:3}(a) the best isocurvature model ($\chi^2 = 80$)
does badly with the COBE
region as well as after the prominent peak. This peak
is fitted quite well by the second acoustic peak while the first acoustic peak
appears as a small shoulder around $\ell \simeq 80$.  

The considerations so far rely on the CMB data only.
However, as is well known, when discussing isocurvature models
it is essential to include also
the large-scale structure (LSS) data. As we will see,
rough measures are already very effective in constraining the models.
Therefore we make use of the
the amplitude of the rms mass fluctuations in
an $8h^{-1}$ Mpc sphere only,
denoted as $\sigma_8$, which the LSS data restricts to the range
$0.43 < \sigma_8\Omega_m^{0.56} < 0.70$ \cite{LSS}.
The contours of $\sigma_8\Omega_m^{0.56}$ are shown in Fig.~\ref{fig:1}(b).
Apart from the upper left corner
of the $(\Omega_m,\Omega_\Lambda)$ plane,
the best-fit closed models $\phantom{x}$ appear to give
a far too large $\sigma_8\Omega_m^{0.56} \gsim 1.5$.
This is natural, since we need a large $\niso$
to do away with the first peak
(``the isocurvature shoulder'') at $\ell \simeq 60 \ldots 100$
and to get enough
power at higher multipoles. A large $\niso$ evidently leads to a large
$\sigma_8$. To compensate for this, one would require a small $\Omega_m$.
We have checked that the smaller $\Omega_m$ we
have, the larger $\niso$ is allowed for by the LSS constraint.
In particular, the upper left corner
closed models in Fig.~\ref{fig:1}b obey
the LSS constraint, although they have a rather large
spectral index $\niso \simeq 3.1$.

On the other hand, the best-fit open models tend to have a slightly
too small $\sigma_8\Omega_m^{0.56}$. These models have a relatively
small
$\niso \lesssim 2.1$, for the following reasons. (1) Since these models fit the
first isocurvature peak to the $\ell \simeq 200$ peak in the data,
they do not need
a large $\niso$ to eliminate this first peak. (2) The smaller scales do not need
as large a boost from $\niso$, since power is provided by the second peak
where the data requires it. Because of this smaller $\niso$ these
models fit the COBE region better.

We have repeated the analysis of minimizing $\chi^2$ but now
with the LSS constraint. As
one might expect, this eliminates most of the best-fit closed models,
leaving only those with a small $\Omega_m$ and a large $\Omega_\Lambda$;
see the upper left corner of Fig.~\ref{fig:2}(a).
The reason for this shifting
of the best-fit closed-model region  to the opposite corner
in the $(\Omega_m,\Omega_\Lambda)$ plane
is easy to understand. Large $\niso$ leads to a large $\sigma_8$, and
hence the prior $0.43 < \sigma_8\Omega_m^{0.56} < 0.70$
requires $\Omega_m$ to be small, which in turn implies
a large $\Omega_\Lambda$ in order to adjust the peak position.

After imposing the LSS constraint, the best-fit model is no longer a closed one
but an open model at the corner of the parameter space with
$\omega_b = 0.001$ and $\Omega_\Lambda = -1.00$.
This fit has $\chi^2 = 103$ and
$(\niso,\Omega_m,\Omega_\Lambda,\omega_b,\omega_c)
= (2.05, 0.71, -1.00, 0.001, 0.16)$. Fig.~\ref{fig:3}(a) shows that
the first acoustic peak at $\ell \simeq 170$ is too low to fit the data.
It is clear that the fit would further improve
 if one allowed for even smaller $\omega_b$ and $\Omega_\Lambda$.  However,
such a small $\omega_b$ is in clear conflict with
big bang nucleosynthesis (BBN).
There is some debate in the BBN community\cite{BBN} on how small an
$\omega_b$ could be acceptable.  After imposing a very conservative lower limit,
$\omega_b \geq 0.003$, our best-fit open model is already significantly worse
than the best-fit closed models.  Moreover, the best-fit open models have a
very small, even a negative, $\Omega_\Lambda$.  This region of the
$(\Omega_m,\Omega_\Lambda)$ plane is disfavored by the observed
supernova redshift-distance relationship\cite{SN}.

Thus we conclude that the best candidates for pure isocurvature models
are the remaining best-fit closed models.
These models satisfy the LSS constraint and have an acceptable
$\omega_b$.  They lie in the region of small $\Omega_m$
and large $\Omega_\Lambda$.  We scanned this region with a finer grid.
The resulting best-$\chi^2$ contours
in the $(\Omega_m,\Omega_\Lambda)$ plane are shown in Fig.~\ref{fig:2}(b)
along with the baryon density of these models.
The best
``physically acceptable'' isocurvature fit has
$(\niso,\Omega_m,\Omega_\Lambda,\omega_b,\omega_c)
= (2.80, 0.12, 0.97, 0.015, 0.074)$.
The fit remains very bad, however, with $\chi^2 = 121$
for $40$ data points and $6$ parameters, to be
compared to $\chi^2 = 44$ of the flat adiabatic reference model.
Because of the high
$\chi^2$ of the best fit, it is unnecessary to consider the LSS spectrum
in a more detailed way.
The badness of the fit is mainly due to the COBE and Boomerang data;
see Fig.~\ref{fig:3}(b).
The COBE contribution to $\chi^2$ is $2.4$ per COBE data point, the
Boomerang contribution is $4.2$ per data point, while the Maxima contribution
remains at $1.7$. The slope of the best-fit model is
the reason for the poor
fit to COBE,
and although the prominent peak in the data is fitted quite well,
the ``flat adiabatic'' peak structure of the second and third peaks in the
Boomerang data leads
to a conflict with the isocurvature peak structure.

As mentioned earlier, the power-law form for the power spectrum
is not necessarily the most natural one in open and closed models
due to the effect of spatial curvature.
The curvature scale in the models studied is comparable to the Hubble
scale, or larger.  Thus its effect is expected
to be reflected in the COBE region of the power spectrum, but not in
the Boomerang/Maxima region.  To assess the significance of this problem,
we repeated our analysis without the COBE data points.  The results
remained essentially unchanged.
Without the 8 COBE points we got $\chi^2 = 70$ for the best-fit model,
$\chi^2 = 91$ for the best-fit with LSS constraint, $\chi^2 = 89$ for
the best physically acceptable fit from the refined grid, and
$\chi^2 = 40$ for the adiabatic reference model.
Hence the Boomerang data alone are sufficient
to rule out pure isocurvature models
and our conclusions do not depend on the question
of the effect of spatial curvature on the power spectrum.

Actually, since the main discriminant is the relative positions
of the three peaks in the Boomerang data, which show an
``adiabatic'' instead of an ``isocurvature'' pattern, our conclusion
should be independent of the shape of the primordial power spectrum
as long as the observed peaks are indeed due to acoustic
oscillations and do not represent features of the
primordial power spectrum itself.

\section{summary}
We have surveyed a large space of  parameters for pure
isocurvature models, and allowed for both open and closed universes,
to find out whether
there are any pure isocurvature models that fit the current CMB data better
than or at least equally as well as the flat adiabatic model.  There
are none.
We conclude that, even if one
ignores the high-$z$ supernova data, pure
isocurvature CDM models, including the ones with a heavily tilted spectrum,
are completely ruled out by the present CMB and LSS data.
Incidentally, the isocurvature models do not do too badly with the Maxima-1
data.  The main CMB problems are with the COBE and the Boomerang data.  To have
sufficient smaller-scale power, and to suppress the first peak and boost the
second peak in the closed models, a large blue tilt is needed.  This leads to
a slope in the Sachs--Wolfe region and reduces
the largest-scale power below the level observed by COBE.  
The most significant problem, however, is with the Boomerang data.
Boomerang shows a second and a third peak with a spacing that corresponds to a
flat universe, whereas the position of the first peak in the data
cannot be fitted by flat isocurvature models.

\section*{Acknowledgments}

This work was supported by the
Academy of Finland  under the contracts 101-35224 and 47213.
We thank Alessandro Melchiorri for a useful communication,
Elina Sihvola for technical
help, and the Center for Scientific Computing (Finland) for computational
resources. We acknowledge the use of the
Code for Anisotropies in the Microwave Background (CAMB)
by Antony Lewis and Anthony Challinor.

\end{document}